\documentclass[prb,onecolumn,12pt]{revtex4}
\usepackage{epsfig}
\usepackage{amssymb}
\usepackage{threeparttable}
\usepackage{graphicx}
\usepackage{longtable}

\begin{document}
\title{Towards intrinsic phonon transport in single-layer MoS$_2$}
\author{Bo Peng$^1$, Hao Zhang$^{1,*}$, Hezhu Shao$^2$, Yuanfeng Xu$^1$, Xiangchao Zhang$^1$ and Heyuan Zhu$^1$}
\affiliation{$^1$Shanghai Ultra-precision Optical Manufacturing Engineering Center, Department of Optical Science and Engineering, Fudan University, Shanghai 200433, China\\
$^2$Ningbo Institute of Materials Technology and Engineering, Chinese Academy of Sciences, Ningbo 315201, China}

\begin{abstract}
The intrinsic lattice thermal conductivity of MoS$_2$ is an important aspect in the design of MoS$_2$-based nanoelectronic devices. We investigate the lattice dynamics properties of MoS$_2$ by first principles calculations. The intrinsic thermal conductivity of single-layer MoS$_2$ is calculated using the Boltzmann transport equation for phonons. The obtained thermal conductivity agrees well with the measurements. The contributions of acoustic and optical phonons to the lattice thermal conductivity are evaluated. The size dependence of thermal conductivity is investigated as well.
\end{abstract}

\maketitle

\section{INTRODUCTION}

Two-dimensional (2D) materials such as graphene and single-layer (SL) transition metal dichalcogenides MX$_2$ (M = Mo, W; X = S, Se, or Te) have drawn considerable interest in recent years due to their potential for integration into next-generation electronic and energy conversion devices \cite{C4NR01600A,Novoselov2012,Klinovaja2013}. SL MoS$_2$ consists of a Mo atom layer sandwiched between two S atom layers, connected by covalent bonds. In the SL form, MoS$_2$ has a direct bandgap of 1.9 eV, and switches to an indirect bandgap of 1.6 eV in double-layer MoS$_2$. The bandgap shrinks when more layers added, and drops to 1.3 eV in the bulk form as a layered semiconductor, where the adjacent layers are connected by the van der Waals force \cite{Mak2010}. Compared to gapless graphene, the extraordinary electronic structure of SL MoS$_2$ may lead to many potential applications such as field-effect transistors \cite{doi:10.1021/nn202852j,Sarkar2014}, optoelectronic devices \cite{Wang2012}, and spin-valley devices \cite{Zeng2012,Xiao2012}. 

Since SL MoS$_2$ is considered to play an important role in the next-generation nanoelectronic devices, it is necessary to investigate the thermal properties of MoS$_2$. High-performance electronic devices strongly depend on high thermal conductivity for highly efficient heat dissipation, while low thermal conductivity is preferred in thermoelectric application. Recently, experimentally measured thermal conductivity $\kappa$ of bulk and SL MoS$_2$ are 110$\pm$20 W/mK \cite{Liu2014} and 84$\pm$17 W/mK \cite{Zhang2015a}, respectively, since phonons are more sensitive to surface disorder with decreasing layers \cite{Jo2014}. In addition, theoretically calculated values of $\kappa$ based on Boltzmann transport equation (BTE) \cite{Li2013a,Gu2014,Cepellotti2015}, Molecular-dynamics (MD) simulations \cite{Liu2013}, the Klemens model \cite{Cai2014,Wei2014,Su2015}, present great disagreement even by orders of magnitude. Therefore, a precise calculation of $\kappa$ of SL MoS$_2$ and clear explanations are needed to clarify such disagreements.

According to the kinetic theory \cite{Yang2013}, accurate prediction of $\kappa$ requires the precise calculation of the distribution of phonon mean free paths (MFPs). Liu $et. al.$ have measured the distribution of phonon MFPs of MoS$_2$ and suggest that approximately 25\% of heat is carried by phonons with MFP larger than 2 $\mu$m \cite{Liu2014}. However, the reported MD simulations predicted the longest MFP to be 5.2 nm \cite{Liu2013}, while some former theoretical calculations based on the Klemens' expression obtained the longest MFP in the scale of 10-200 nm \cite{Cai2014,Wei2014,Su2015}, which is much lower than the measured value. As a result, the corresponding estimated $\kappa$ is from 1.35 W/mK to 29.2 W/mK, which is lower than the measured 84$\pm$17 W/mK as well.

In this paper, we obtain phonon properties from lattice dynamics calculations. The thermal conductivity of isotopically pure and naturally occurring SL MoS$_2$ are calculated using an iterative solution of the BTE for phonons. The calculated thermal conductivity agrees well with the measurements. Accurate relaxation time and MFP of MoS$_2$ is obtained. The calculated MFP distribution of MoS$_2$ is in consistency to the experimental results. The MFP in the small-grain limit is also investigated when the nanostructuring induced phonon scattering dominates. The role of boundary scattering in MoS$_2$ nanowires is examined as well.

\section{METHODOLOGY}

The in-plane $\kappa$ can be calculated as a sum of contribution of all the phonon modes $\lambda$, which comprises both a phonon branch index $p$ and a wave vector $\textbf{q}$,
\begin{equation}\label{kappa-eq}
\kappa=\frac{1}{N_K}\sum_{\lambda}C_{\lambda}v_{\lambda\alpha}^2\tau_{\lambda\alpha},
\end{equation}

\noindent where $N_K$ is the number of sampling points, $C_{\lambda}$ is the heat capacity per mode, $v_{\lambda\alpha}$ and $\tau_{\lambda\alpha}$ are the group velocity and relaxation time of mode $\lambda$ along $\alpha$ direction.

The phonon properties ($C_{\lambda}$, $v_{\lambda\alpha}$, $\tau_{\lambda\alpha}$) in Eq.~(\ref{kappa-eq}) can be obtained through lattice dynamics calculations and an iterative solution of the phonon BTE. The total lattice energy can be expanded by a Taylor series with respect to atomic displacements, which includes harmonic and (mainly third-order) anharmonic terms \cite{Ziman}. The harmonic and anharmonic interatomic force constants (IFCs) can be obtained from harmonic and anharmonic terms respectively. Using the harmonic IFCs, the phonon dispersion relation can be obtained, which determines the group velocity $v_{\lambda\alpha}$ and specific heat $C_{\lambda}$. The contribution to $v_{\lambda\alpha}$ and $C_{\lambda}$ from anharmonic terms is usually negligible, and can be thus neglected in the corresponding calculations. However, the third-order term plays an important role in calculating the three-phonon scattering rate, which is the inverse of $\tau_{\lambda\alpha}$.

All the calculations are performed using the Vienna $Ab$-$initio$ Simulation Package (VASP) based on the density functional theory (DFT) \cite{Kresse1996}. We use the projected augmented wave (PAW) method, and the generalized gradient approximation (GGA) in the Perdew-Burke-Ernzerhof (PBE) parametrization for the exchange-correlation functional. A cutoff of 500 eV is used for the plane-wave expansion. A 15$\times$15$\times$1 \textbf{k}-mesh is used for the unit cell during structural relaxation. The structures are relaxed until the energy differences are converged within 10$^{-8}$ eV, with a Hellman-Feynman
force convergence threshold of 10$^{-4}$ eV/\AA. We maintain the interlayer vacuum spacing larger than 10 \AA\ to eliminate interactions between adjacent supercells. 

The harmonic IFCs are obtained by density functional perturbation theory (DFPT) using the supercell approach, which calculates the dynamical matrix through the linear response of electron density \cite{DFPT}. A 5$\times$5$\times$1 supercell with 5$\times$5$\times$1 \textbf{q}-mesh is used to calculate the dynamical matrix, which is the Fourier transform of the real-space harmonic IFCs $\Phi_{\alpha\beta}$, and can be given by
\begin{equation}
\label{dynamical matrix}
\widetilde{D}_{\alpha\beta,IJ}(\textbf{q})=\sum_{I}\frac{1}{\sqrt{M_IM_J}}\Phi_{\alpha\beta}(\textbf{R}_I)\exp(-i\textbf{q}\cdot \textbf{R}_I),
\end{equation}

\noindent where $\alpha$ and $\beta$ are the Cartesian indices, and $I/J$ is the $I$/$J$-th atom. The phonon frequencies and eigenvectors can be directly obtained by the solution to the eigenvalue equation
\begin{equation}
\label{eigenvalue}
\sum_{\beta,J}\widetilde{D}_{\alpha\beta,IJ}(\textbf{q})e_{\beta,J}(\textbf{q})=\omega^2e_{\alpha,I}(\textbf{q}).
\end{equation}

The anharmonic third order IFCs are calculated using a  supercell-based, finite-difference method \cite{Li2012},
\begin{equation}
\label{third-order IFCs}
\Phi_{\alpha\beta\gamma}^{IJK}=\frac{\partial E^3}{\partial \textbf{R}_I^{\alpha}\partial \textbf{R}_J^{\beta}\partial \textbf{R}_K^{\gamma}}.
\end{equation}

\noindent The same 5$\times$5$\times$1 supercell and 5$\times$5$\times$1 \textbf{q}-mesh are used to obtain the anharmonic IFCs. A well-converged interaction range of 4.2 \AA\ is considered herein, which includes forth-nearest-neighbor atoms.

The lattice thermal conductivity $\kappa$ can be calculated iteratively using the ShengBTE code, which is completely parameter-free and based only on the information of the chemical structure \cite{Omini1996,ShengBTE,Li2012a,Li2012}. A discretizationa of the Brillouin zone (BZ) into a $\Gamma$-centered regular grid of 90$\times$90$\times$1 $\textbf{q}$ points is introduced.

\section{RESULTS AND DISCUSSION}

\subsection{Lattice dynamics properties}

\begin{figure}
\centering
\includegraphics[width=\linewidth]{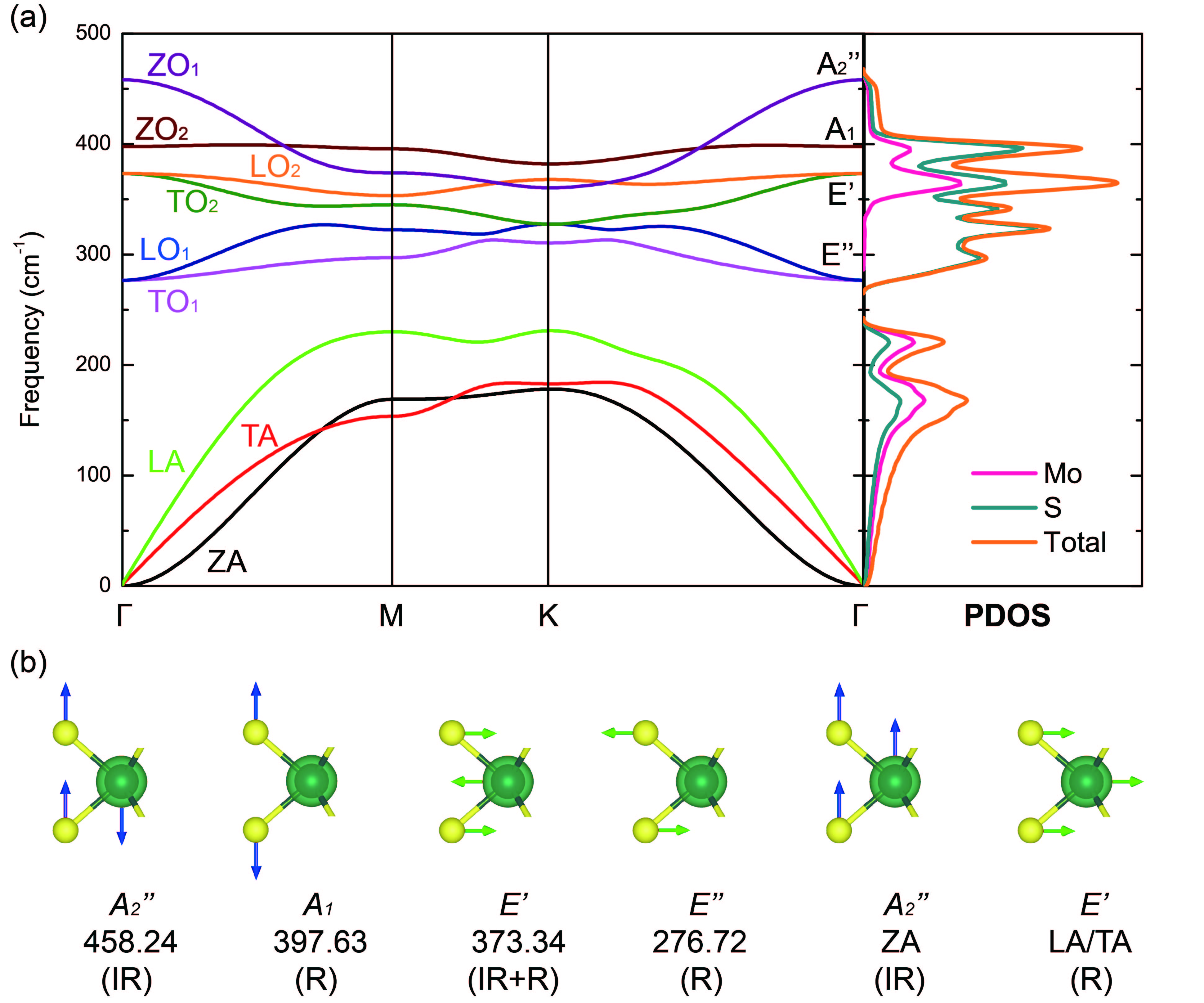}
\caption{(a) Phonon dispersion and DOS of SL MoS$_2$. (b) Oscillation patterns for phonon modes at the $\Gamma$ point.}
\label{fig1} 
\end{figure}

The calculated phonon band structure and phonon density of states (DOS) of SL MoS$_2$ are shown in Fig.~\ref{fig1}(a), in which the lattice vibration modes are characterized by three acoustic [longitudinal acoustic (LA) and transverse acoustic (TA) branches in the basal plane, and flexural acoustic (ZA) branch perpendicular to the basal plane], and six optical branches. The LA and TA branches are linear and the ZA branch is quadratic in the vicinity of the $\Gamma$ point, which is due to the low lattice dimensionality \cite{Huang2015}. The calculated bandgap between the acoustic and optical branches is about 46 cm$^{-1}$, which is in good agreement with other previous theoretical results \cite{Gu2014,Cai2014}.

According to the group-theoretical analysis \cite{C4CS00282B}, since the SL MoS$_2$ belongs to the $D_{3h}$ point group symmetry, the optical lattice-vibration modes at $\Gamma$ can be thus decomposed as, 
\begin{equation}
\Gamma_{optical}^{SL}=A_2''(\mathrm{IR})+A_1'(\mathrm{R})+E'(\mathrm{IR+R})+E''(\mathrm{R}),
\label{opt-mode}
\end{equation}

\noindent where IR and R denote infrared- and Raman-active modes respectively. The oscillation patterns of the optical modes are shown in Fig.~\ref{fig1}(b). Obviously, $A_2''$ and $A_1$ modes are out-of-plane vibration modes, while $E'$ and $E''$ are in-plane vibration modes. Table~\ref{mode} lists the calculated frequencies of the four optical phonon modes at the $\Gamma$ point compared to experimental values for MoS$_2$. The calculated phonon frequencies are in agreement with the experimental results, and the discrepancy is less than 4\%. The LO/TO splitting is very small and can be neglected here \cite{Cai2014}.

\begin{table}
\centering
\caption{Phonon frequencies of the four optical modes at the $\Gamma$ point compared to experimental data (in the units of cm$^{-1}$).}
\begin{tabular}{ccccc}
\hline
 & $A_2''$ & $A_1$ & $E'$ & $E''$ \\
\hline
This work & 458.24 & 397.63 & 373.34 & 276.72 \\
Experimental & 470 \cite{JimenezSandoval1991} & 402.4 \cite{Rice2013} & 383.5 \cite{JimenezSandoval1991} & 287 \cite{Wieting1971} \\
\hline
\end{tabular}
\label{mode}
\end{table}

The Debye temperature $\theta_D$ can be calculated with the highest frequency of normal mode vibration (Debye frequency) $\nu_m$,
\begin{equation}
\theta_D=h\nu_m/k_B
\end{equation}

\noindent where $h$ is the Planck constant, and $k_B$ is the Boltzmann constant. The calculated Debye temperatures $\theta_D$ for MoS$_2$ is 262.3 K, which are in good agreement with previous results, $i.e.$ 260-320 K for MoS$_2$ estimated from specific-heat measurement \cite{debye-260-MoS2}. The Debye temperature reflects the magnitude of sound velocity. Higher Debye temperature means increased phonon velocities and increased acoustic-phonon frequencies, which suppress phonon-phonon scattering by decreasing phonon populations \cite{Lindsay2013,pb2}.

\begin{figure}
\centering
\includegraphics[width=0.85\linewidth]{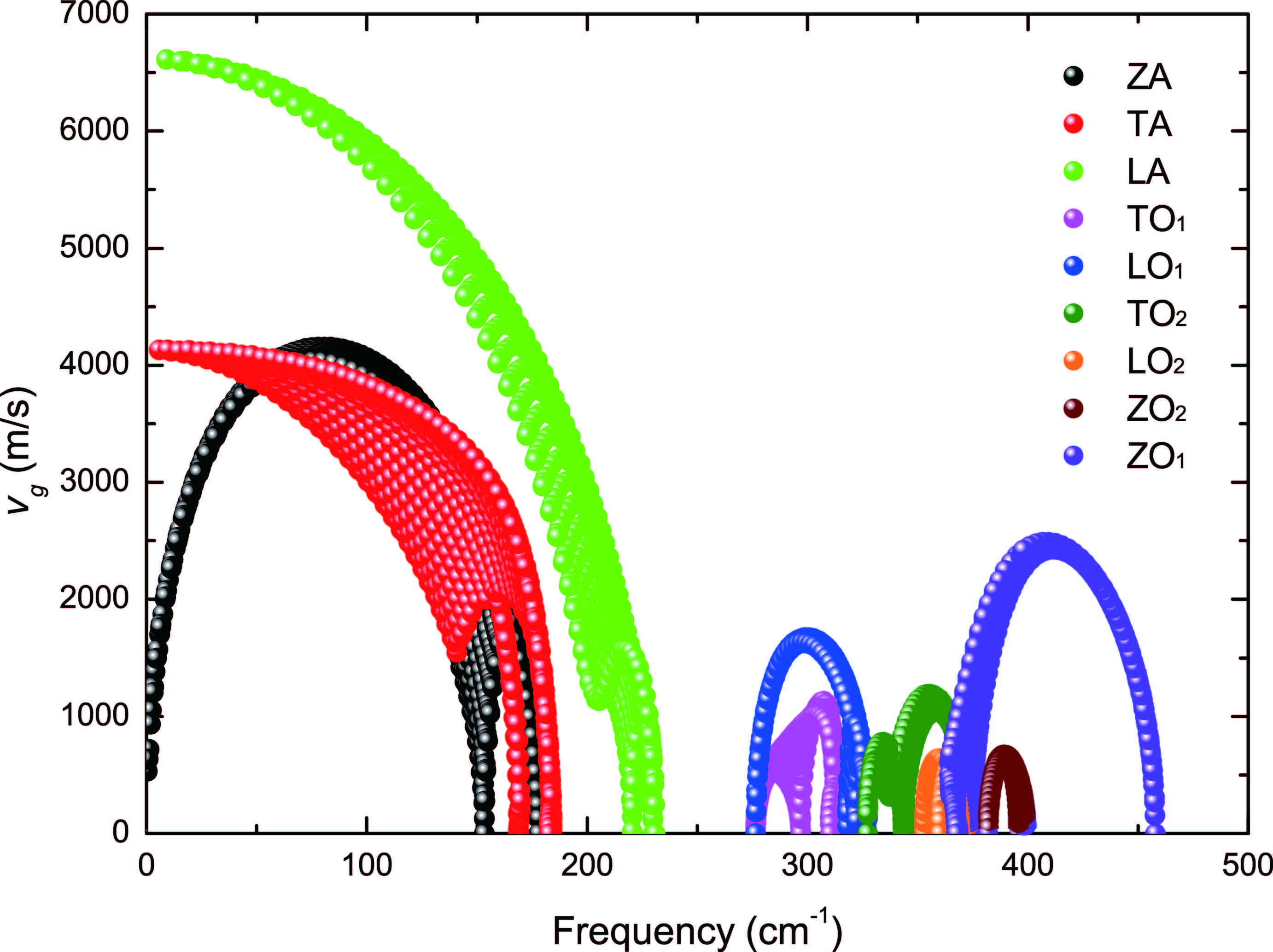}
\caption{Calculated phonon group velocities of MoS$_2$ in the irreducible BZ.}
\label{vg} 
\end{figure}

The group velocities $v_g$ in the irreducible BZ are shown in Fig~\ref{vg}, which indicate weak anisotropy at high frequencies within the entire BZ. The sound velocities in long-wavelength limit are about 4,093 and 6,549 m/s for the TA and LA modes respectively, which are comparable to 5,400-8,800 m/s in silicene \cite{Li2013}, and 4,000-8,000 m/s in blue phosphorene \cite{Jain2015}. The group velocities of optical phonons are less than 3000 m/s, which are much smaller than acoustic phonons.

\subsection{Intrinsic thermal conductivity}

\begin{figure}
\centering
\includegraphics[width=0.85\linewidth]{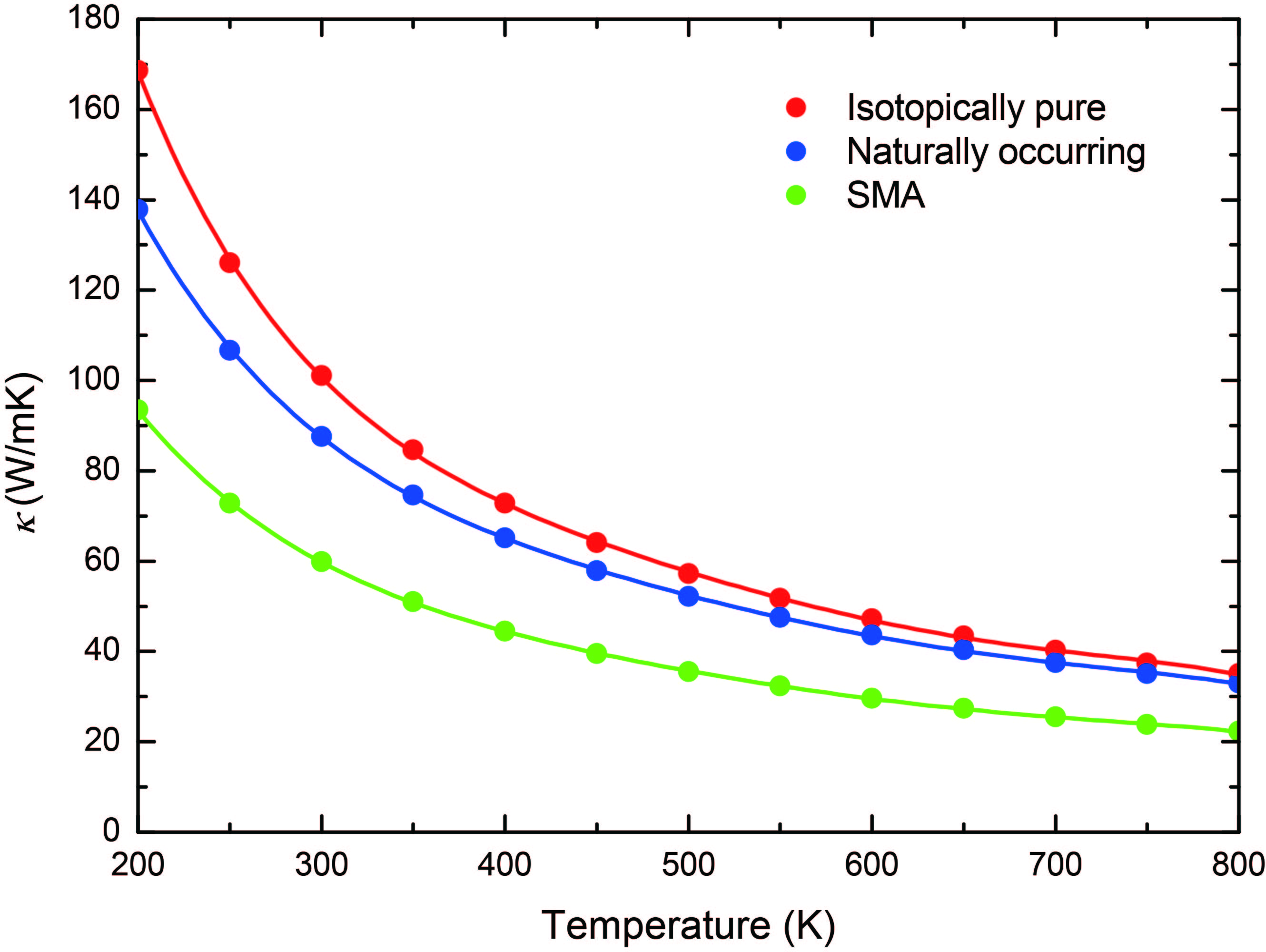}
\caption{Lattice thermal conductivity of isotopically pure and naturally occurring SL MoS$_2$, in comparison with the results of naturally occurring MoS$_2$ calculated from the single-mode relaxation time approximation (SMA).}
\label{kappa} 
\end{figure}

Fig.~\ref{kappa} presents the lattice thermal conductivity $\kappa$ of isotopically pure and naturally occurring MoS$_2$ using the iterative solution of the BTE. The intrinsic $\kappa_{pure}$ of MoS$_2$ is 101.0 W/mK, while for naturally occurring isotope concentrations, the $\kappa_{iso}$ is 87.6 W/mK, which agrees well with the measured value of 84$\pm$17 W/mK \cite{Zhang2015a}.

Naturally occurring Mo consists of 14.84\% $^{92}$Mo, 9.25\% $^{94}$Mo, 15.92\% $^{95}$Mo, 16.68\% $^{96}$Mo, 9.55\% $^{97}$Mo, 24.13\% $^{98}$Mo, and 9.63\% $^{100}$Mo, while naturally occurring S consists of 95.02\% $^{32}$S, 0.75\% $^{33}$S, 4.21\% $^{34}$S, and 0.02\% $^{36}$S.

The introduction of isotope disorder can be used to modulate 10\% of the thermal conductivity in isotopically pure MoS$_2$ at temperatures below 262.3 K. The $\kappa$ becomes less sensitive to isotopes at high temperatures, since the Umklapp processes (U processes) become frequent enough and drive the thermal conductivity at temperatures well above $\theta_{D}$ \cite{Ziman,Grimvall}.

The comparison between $\kappa$ of SL MoS$_2$ with naturally occurring isotope concentrations from the single-mode relaxation time approximation (SMA) and the exact solution of the BTE as a function of temperature is also shown in Fig.~\ref{kappa}. The SMA assumes that individual phonon mode is excited independently, which has no memory of the initial phonon distribution, therefore the SMA is inadequate to describe the momentum-conserving character of the Normal processes (N processes) and it works well only if the U processes dominate \cite{Fugallo2013}. With increasing temperature, the difference between the two approaches declines. Our results indicate the N processes play an important role in thermal transport at low temperatures, and the U processes become more important with increasing temperature.

\begin{table}
\centering
\caption{Contribution of different phonon branches (ZA, TA, LA, and all optical) towards the total thermal conductivity in MoS$_2$, in comparison with graphene and stanene.}
\begin{tabular}{ccccc}
\hline
 & ZA (\%) & TA (\%) & LA (\%) & Optical (\%) \\
\hline
MoS$_2$ & 29.1 & 30.4 & 39.1 & 1.4 \\
graphene \cite{Lindsay2014} & 76 & 15 & 8 & 1 \\
stanene \cite{pb3} & 13.5 & 26.9 & 57.5 & 2.1 \\
\hline
\end{tabular}
\label{contribution}
\end{table}

The contributions of different phonon branches to $\kappa_{iso}$ are listed in Table~\ref{contribution}, in comparison with graphene and stanene. It has been reported that the large contribution of ZA phonons to the $\kappa$ of graphene is due to a symmetry selection rule in one-atom-thick materials, which strongly restricts anharmonic phonon-phonon scattering of the ZA phonons \cite{Lindsay2010}, while the buckled structure in stanene breaks out the out-of-plane symmetry. As for MoS$_2$, ZA and TA phonons contribute almost equally to $\kappa$, while the LA contribution to $\kappa$ is a bit larger than that from the other two phonon modes. Considering the significant difference in the contribution of each acoustic phonon mode to the total thermal conductivity in these materials, it is worthwhile to perform a detailed investigation of the scattering mechanism in MoS$_2$.

\begin{figure}
\centering
\includegraphics[width=0.85\linewidth]{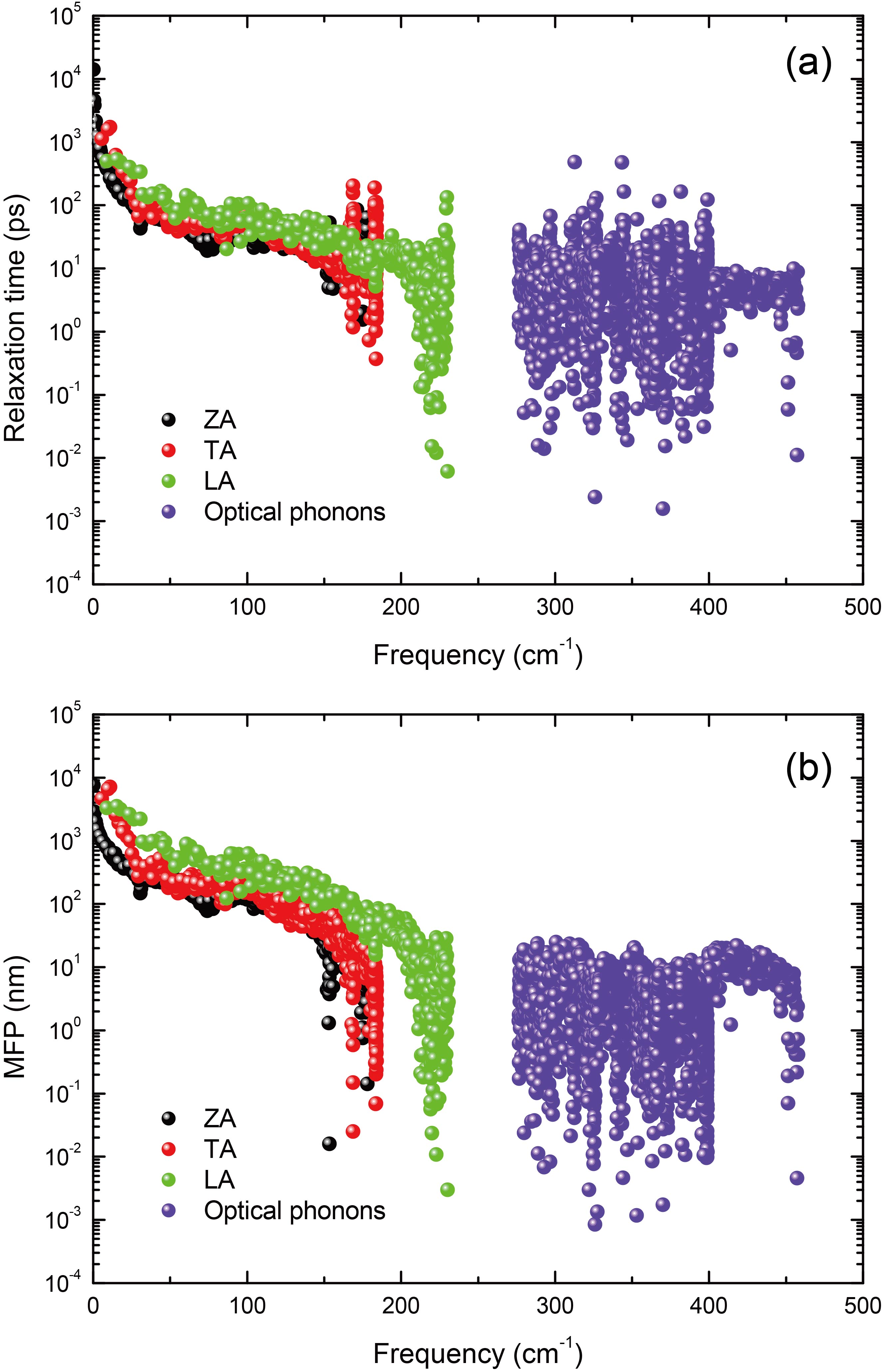}
\caption{Frequency dependence of (a) phonon relaxation time and (b) MFP at 300 K.}
\label{time} 
\end{figure}

We extract the frequency-dependent relaxation time of SL MoS$_2$ in Fig.~\ref{time}(a). At frequencies below 20 cm$^{-1}$, the relaxation time of ZA phonons are larger than TA phonons, while LA phonons have much short relaxation time. The relaxation time of three acoustic phonon modes is comparable with each other at higher frequencies. The frequency-dependent MFP of each phonon branch is calculated by $l_{\lambda}=v_{\lambda}\tau_{\lambda}$, as shown in Fig.~\ref{time}(b). The MFP of LA phonons is slightly larger than the other two branches at frequencies above $\sim$25 cm$^{-1}$ due to larger group velocities. Thus the LA phonons contribute the largest part to the total $\kappa$.

The calculated longest relaxation time of acoustic phonons ($\sim$10$^4$ ps) is two to three orders of magnitude larger than the previous work ($\sim$20-500 ps) \cite{Cai2014,Su2015} based on the Klemens' expressions \cite{Klemens1994}. As a result, the longest MFP in Ref. \cite{Cai2014} and \cite{Su2015} (18.1 nm and 221.4 nm, respectively) is much smaller than measured value of $>$2 $\mu$m \cite{Liu2014}. In this work, the calculated MFP in Fig~\ref{time}(b) is in consistency to the experimental result.

\subsection{Size dependence of $\kappa$}

\begin{figure}
\centering
\includegraphics[width=0.85\linewidth]{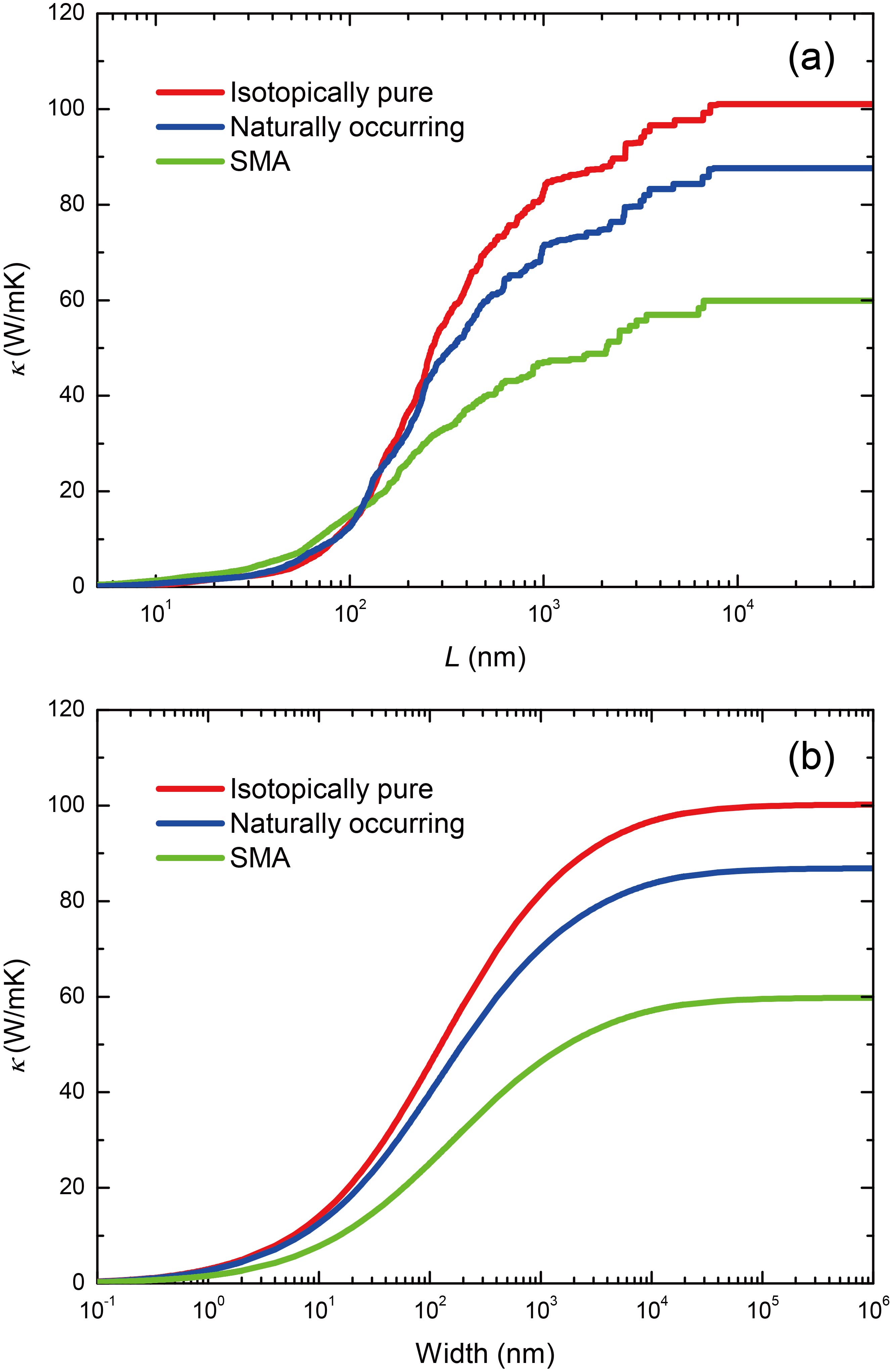}
\caption{(a) Cumulative thermal conductivity as a function of phonon MFP of isotopically pure and naturally occurring SL MoS$_2$ at 300 K, in comparison with the results of naturally occurring MoS$_2$ calculated from the single-mode relaxation time approximation (SMA). (b) Thermal conductivities of MoS$_2$ nanowires along
[100] direction as a function of width.}
\label{MFP} 
\end{figure}

The MFP distribution plays an important role in describing the behaviour of phonon transport within a sample. Generally, when the characteristic length $L$ is smaller than the phonon MFP, $i.e.$ $L<l_{\lambda}$, phonons move ballistically without collisions, and phonon transport is in the ballistic regime, in which the thermal conductivity is smaller than the prediction by Fourier's law. When $L>l_{\lambda}$, phonon transport changes from the ballistic regime to the diffusive regime, in which Fourier's law is valid. The typical procedure to analyze the MFP distribution is to calculate the cumulative thermal conductivity as a function of MFP \cite{Yang2013,Fugallo2014}. 

The cumulative thermal conductivity with MFPs below $L$ can be calculated by the following expression,
\begin{equation}
\kappa(L)=\frac{1}{V}\sum_{\lambda}^{l_{\lambda}<L}C_{\lambda}v_{\alpha}^2\tau_{\lambda\alpha}.
\end{equation}

\noindent The cumulative thermal conductivity with respect to $L$ at 300 K is shown in Fig.~\ref{MFP}(a). The accumulation $\kappa(L)$ increases as $L$ increases, until reaching the thermodynamic limit above a length $L_{diff}$, which represents the longest mean free path of the heat carriers. The $L_{diff}$ of isotopically pure and naturally occurring MoS$_2$ are 7.5 $\mu$m and 7.2 $\mu$m, respectively, which are in consistency to the experimental result \cite{Liu2014}. The MFP distribution also suggests that, to reduce the lattice thermal conductivity of SL MoS$_2$, a sample with a characteristic length less than $\sim$7 $\mu$m is required.

Furthermore, when the size of sample gets smaller, the nanostructuring-induced phonon scattering becomes dominant over the three-phonon scattering, and the small-grain-limit reduced $\kappa$ becomes proportional to a constant value $l_{SG}$ \cite{ShengBTE}. We calculate the ratio of the thermal conductivity to the thermal conductivity per unit of MFP in the small-grain limit; the $l_{SG}$ is found to be 131.5 nm and 114.1 nm for isotopically pure and naturally occurring MoS$_2$ at 300 K. This quantity is crucial for the thermal design to modulate the thermal conductivity in the small-grain limit, for example nanowires.

In a nanowire system, phonons with long MFPs will be strongly scattered by the boundary. As a result, the contribution of these phonons to $\kappa$ will be limited. In MoS$_2$ nanowires, $\kappa$ decreases with decreasing width, and drops to about half the maximum $\kappa$ in the thermodynamic limit at widths about 130 nm and 126 nm for isotopically pure and naturally occurring MoS$_2$, as shown in Fig.~\ref{MFP}(b). Our result indicates that the lattice thermal conductivity of MoS$_2$ is sensitive to boundary scattering, and can be further reduced in
nanostructures for engineering thermal transport in MoS$_2$.

\section{Conclusion}

We calculate the lattice thermal conductivity $\kappa$ of SL MoS$_2$ using first-principle calculation and an iterative solution of the BTE for phonons. The introduction of isotopes leads to a 10\% reduction of $\kappa$. The intrinsic relaxation time and the distribution of phonon MFPs are investigated in detail. The diffusion-limited MFP SL MoS$_2$ is larger than 7 $\mu$m at 300 K. The size dependence of thermal conductivity is investigated as well for the purpose of designing nanostructures. Our work provides a fundamental understanding of phonon transport in SL MoS$_2$ to predict the thermal performance of MoS$_2$-based potential devices.

\section*{Acknowledgement}
This work is supported by the National Natural Science Foundation of China under Grants No. 11374063 and 11404348, and the National Basic Research Program of China (973 Program) under Grants No. 2013CAB01505.

\providecommand{\WileyBibTextsc}{}
\let\textsc\WileyBibTextsc
\providecommand{\othercit}{}
\providecommand{\jr}[1]{#1}
\providecommand{\etal}{~et~al.}

\end{document}